\documentclass[showpacs,showkeys,aps,prb,twocolumn]{revtex4}
\usepackage[english]{babel}
\usepackage{epsfig}
\usepackage{graphicx}
\usepackage{amsmath}
\usepackage{amssymb}
\usepackage{subfigure}
\usepackage{bm}
\usepackage{subfigure}

\begin{document}

\title{Phase separation in doped systems with
spin-state transitions}

\author{A.~O.~Sboychakov}

\affiliation{Institute for Theoretical and Applied Electrodynamics,
Russian Academy of Sciences, Izhorskaya Str. 13, Moscow, 125412
Russia}

\author{K.~I.~Kugel~\cite{affUK}}

\affiliation{Institute for Theoretical and Applied Electrodynamics,
Russian Academy of Sciences, Izhorskaya Str. 13, Moscow, 125412
Russia}

\author{A.~L.~Rakhmanov~\cite{affUK}}

\affiliation{Institute for Theoretical and Applied Electrodynamics,
Russian Academy of Sciences, Izhorskaya Str. 13, Moscow, 125412
Russia}

\author{D.~I.~Khomskii~\cite{affUK}}

\affiliation{$II.$ Physikalisches Institut, Universit\"at zu
K\"oln, Z\"ulpicher Str. 77, 50937 K\"oln, Germany}

\begin{abstract}

Spin-state transitions, observed in many transition metal
compounds containing Co$^{3+}$ and Fe$^{2+}$, may occur
with the change of temperature, pressure, but also with
doping, in which case the competition of single-site
effects and kinetic energy of doped carriers can favor
a change in the spin state. We consider this situation
in a simple model, formally resembling that used for
manganites in Ref.~\onlinecite{myPRL}.
Based on such a model, we predict the possibility
of a jump-like change in the number of Co$^{3+}$
ions undergoing spin-state transition caused by hole doping.
A tendency to the electronic phase separation within a
wide doping range is demonstrated. Phase diagrams with
the regions of  phase separation are constructed at different
values of the characteristic parameters of the model.
\end{abstract}



\date{\today}

\maketitle

\section{Introduction}\label{Intr}

Interplay of different degrees of freedom and different types of
ordering is a very important ingredient in determining the properties of strongly correlated electron systems. Especially interesting these effects become in doped systems. Typical for this case is the tendency
to phase separation and formation of inhomogeneous states. It can take
different forms: formation of isolated polarons or small clusters
(modification of a particular ordering by doped charge carriers and
the trapping of charge carriers in a distorted region)
or particular textures, e.g. stripes. Such phase separation can play
a very important role in many phenomena, such as colossal
magnetoresistance in manganites~\cite{dagbook}, and probably also in
high-Tc superconductors - although their role in the latter is still
a matter of hot debate.

The most common and best known case is the doping of
antiferromagnetic insulators, with the formation of ferromagnetic
droplets (``ferrons'') or charged antiferromagnetic domain walls
(stripes)~\cite{dagbook,Nag,Kak}. We have recently shown that,
similarly, the interplay of kinetic energy of doped holes with the
orbital structure can give rise to a novel mechanism of phase
separation~\cite{orbPS1,orbPS2}.

A special interesting group of phenomena is met in systems where
the respective ions can exist in different multiplet states. Typical
examples are the compounds containing Co$^{3+}$ (or sometimes
Fe$^{2+}$), which can exist in a low-spin (LS) state with $S$=0
($t_{2g}^6$), intermediate-spin (IS) state $S=1$ ($t_{2g}^5e_g^1$),
and high-spin (HS) state ($t_{2g}^4e_g^2$) with $S=2$, see e.g.
Ref.~\onlinecite{Gooden67}. Close proximity in energy of these
states can lead to a special type of transition (or crossover):
spin-state transition (SST), typical example being
LaCoO$_3$~\cite{Gooden67,Tokura1,Tokura2,Korotin}. Also spin-state ordering is possible~\cite{Doumerc,KhoLow}. Thus, these systems, in addition to quite common charge, orbital, and spin degrees of freedom with the possibility of respective orderings, have an ``extra dimension'': the possibility of spin-state (or, in other words, multiplet) transitions. Correspondingly, if doping of materials like manganites can cause phase separation due to an interplay of the motion (kinetic energy) of doped  holes  with the underlying magnetic and orbital structure, in systems with SST like cobaltites one can expect similar phenomena due to an interplay with the spin state of the matrix. The common mechanisms causing the phase separation manifest themselves in the situation when the particular ordering existing in the system hinders the motion of doped holes. In these cases, it may be favorable to locally modify the type of ordering, facilitating the motion of the hole in such distorted region. Thus, holes can hardly move on an antiferromagnetic background, which was noticed already long ago both for the two-band (double exchange) model~\cite{AndHas,deGennes} and for the single-band (Hubbard)
model \cite{BulKhom67,BulNagKhom}. At the same time a hole moves
freely on the ferromagnetic background. As a result, ferromagnetic
polarons (ferrons) may be formed close to the hole, and the gain in kinetic energy of the latter moving on the ferromagnetic background exceeds the loss of the magnetic energy~\cite{Nag67,Kasuya,BulNagKhom}.

Similarly, certain types of orbital ordering suppress hole motion,
and it may be favorable to modify orbital pattern close to a hole,
forming orbital polaron and facilitating motion of a hole within it
~\cite{orbPS1,KilKha,MiKhoSaw}. For systems with SST such a
role can be played for example by the phenomenon of a spin blockade
\cite{SpinBlock}: if one dopes the material with the Co$^{3+}$ in a low-spin state ($S=0$) by electrons, the ionic state created
could be Co$^{2+}$ in a high-spin state ($S=3/2$). In this case,
it is evident that it is not possible  to interchange the states
Co$^{3+}$ LS and Co$^{2+}$ HS by moving only one electron: one
would end up in the ``wrong" states Co$^{3+}$ and Co$^{2+}$ both in IS states, not in the original states (the hopping of an electron can change the spin of corresponding states only by $\pm 1/2$, whereas the spins of the original states differ by 3/2). As a consequence, an extra electron can only move in a crystal leaving the trace of wrong spin states, which will lead to a confinement and localization of this electron, similar to the case of the usual Hubbard model~\cite{BulNagKhom}.

One can ``repair" this by modifying the spin state in the vicinity
of a charge carrier (electron or hole), and this will finally again lead to a creation of inhomogeneous states and to phase separation. This phenomenon was actually observed in some cobaltites, e.g. in
La$_{1-x}$Sr$_x$CoO$_3$. There are already many indications of phase
separation and formation of inhomogeneous states in this
system~\cite{Rivas,Loshk,Louca,Leighton06,Leighton07},
but probably the most clear evidence comes from the study of very
low doped LaCoO$_3$. Magnetic measurements~\cite{Tokura1} have shown
that at very low doping ($< 1 \% $ of Sr) the moment per doped hole (per Sr) is much bigger than that of only a LS Co$^{4+}$ with $S=1/2$: instead there exist magnetic impurities with unusually large spin $S=5-10$, which signals that each hole is ``dressed" by the magnetic cloud due to the promotion of some of neighboring Co$^{3+}$ ions to a magnetic state. The magnetic scattering, ESR and NMR study of such system~\cite{Podles08} allowed even determining the size and shape of such magnetic clusters formed around doped holes.

It is possible to use different approaches to describe theoretically the phenomenon of phase separation. First of all, it is the direct numerical investigation~\cite{dagbook}. Or one can assume the formation of spin-state polarons, calculate their energy and check whether and at which conditions the formation of such polarons can be energetically favorable. But the most direct way, by which one usually starts, is first to assume the existence of a homogeneous state  and to check for its stability against phase separation. This was the route taken earlier by us for the double exchange model~\cite{KaKhoMo}, for the situation close to a charge ordering~\cite{KaKuKh01}, for two-component model of manganites~\cite{myPRL,myPRB}, or for orbital
ordering~\cite{orbPS1}. If the homogeneous state turns out to be
unstable, then at the second step one can investigate particular
types of inhomogeneous states, which can be formed. In the present
paper, we follow this route for the doped systems with SST.

\section{Spin states of cobalt ions}\label{SpinSt}

Let us list the possible spin states of Co$^{3+}$ and Co$^{4+}$ ions
in a CoO$_6$ octahedron, which is a main building block of
perovskite-like Co-based compounds (we will consider below the hole-doped cobaltites, nominally containing Co$^{3+}$ and Co$^{4+}$). The electron configuration of Co$^{3+}$ ion is $3d^6$. It is well known that in the crystal field of cubic symmetry, a $d$-level with the 5-fold orbital degeneracy is split into a doubly degenerate $e_g$ level and a triply degenerate $t_{2g}$ level. In the octahedral coordination, the $t_{2g}$ level lies below the $e_g$ level. So, a Co$^{3+}$ ion can have three low-energy spin states: low-spin (LS), intermediate-spin (IS), and high-spin (HS) states.

In the LS state ($S=0$), all $t_{2g}$ states are occupied and the
$e_g$ level is empty. In the IS state ($S=1$), there are five
electrons at the $t_{2g}$ level and one $e_g$ electron. In the HS
state ($S=2$), we have four $t_{2g}$ and two $e_g$ electrons. The
corresponding energies of these states are $E_{LS}^{(3+)} = E_0$,
$E_{IS}^{(3+)} = E_0 + \Delta -J_H$, and $E_{HS}^{(3+)} = E_0 +
2\Delta -4J_H$, where $\Delta$ is the energy splitting between
$t_{2g}$ and $e_g$ levels and $J_H$ is the Hund's rule coupling
constant. For the Co$^{4+}$ ion ($3d^5$), there are three similar
low-lying spin states, corresponding to different distributions of
five electrons between $t_{2g}$ and $e_g$ levels. In the LS state
($S=1/2$), there are five electrons at the $t_{2g}$ level and no
$e_g$ electrons. In the IS state ($S=3/2$), we have four $t_{2g}$
electrons  and one $e_g$ electron. In the HS state ($S=5/2$), the
numbers of $t_{2g}$ and $e_g$ electrons are equal to three and two,
respectively. The corresponding energies of these states for the
Co$^{4+}$ ion are $E_{LS}^{(4+)} = E_1$, $E_{IS}^{(4+)} = E_1 +
\Delta -2J_H$, and $E_{HS}^{(4+)} = E_1 + 2\Delta -6J_H$. Here, we
introduced  $E_0$ and $E_1$ as some reference energy values for Co$^{3+}$ and Co$^{4+}$, respectively. As we shall demonstrate below, the results do not depend much on the specific choice of $E_0$ and $E_1$. All aforementioned configurations of Co ions and their energies are summarized in Table~\ref{Table}.

\begin{table}[tbp]
\includegraphics[width=\columnwidth]{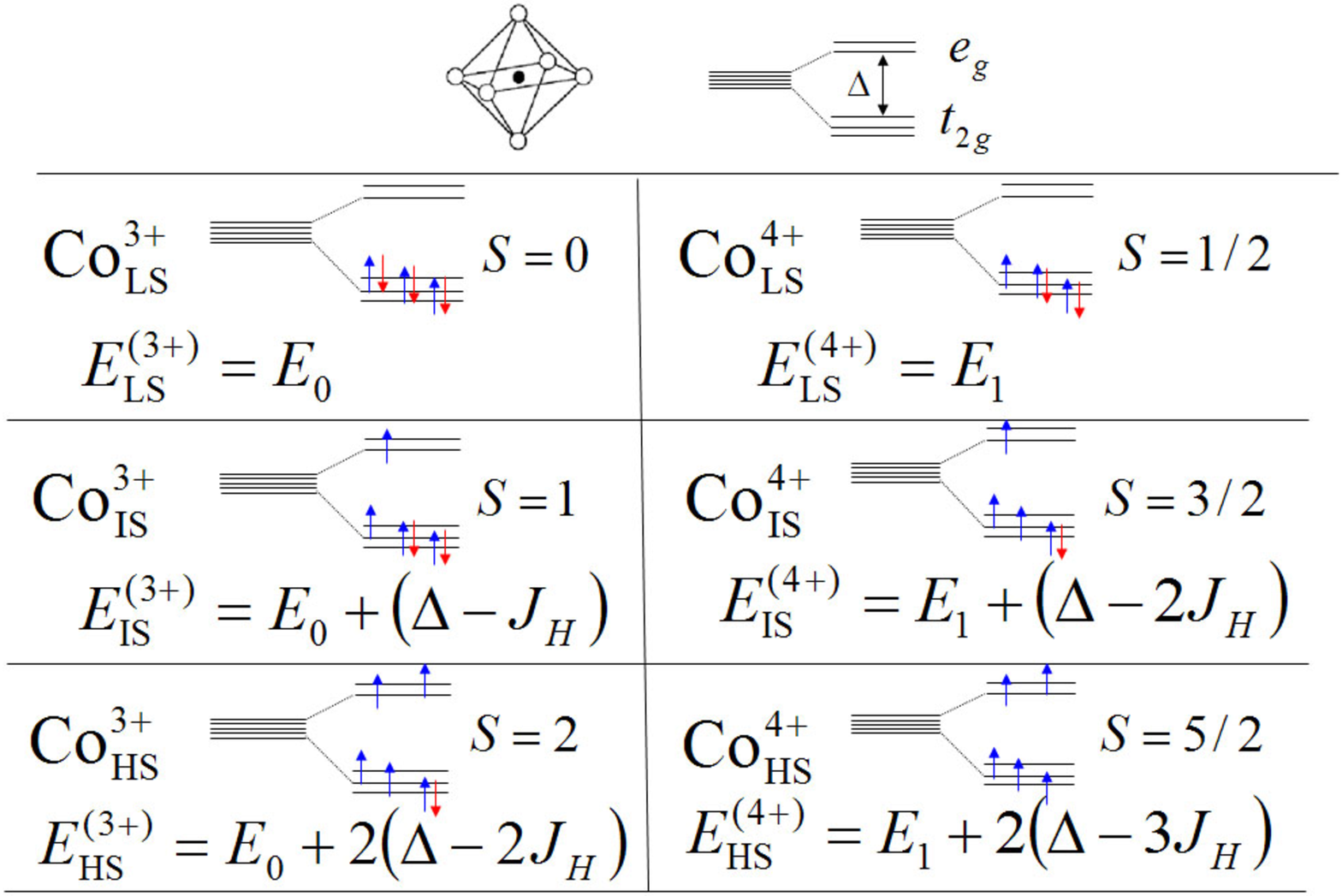}
\caption{(Color online) Possible electron configurations of Co ions
and their energies.
}
 \label{Table}
 \end{table}

The type of the ground state for a separate Co$^{3+}$ or Co$^{4+}$
ion depends on the relationship between  $\Delta$ and $J_H$. It
can be easily seen that at $\Delta > 3J_H$, the LS is the ground
state both for Co$^{3+}$ and Co$^{4+}$. At $2J_H < \Delta < 3J_H$,
Co$^{3+}$ still has the LS ground state, whereas for Co$^{4+}$ the
HS is more favorable. Eventually, at $\Delta < 2J_H$, the HS state
is the lowest in energy for both ions. Hence, for isolated cobalt
ions, the IS ground state does not appear.

The situation becomes more complicated if there exists a charge
transfer between cobalt ions. First, note that the hopping
integrals between the $t_{2g}$ states in cobaltites are as a rule
much smaller than for the $e_g$ states. In the treatment below, we ignore the $t_{2g}-t_{2g}$ hopping and take into account only the hopping of $e_g$ electrons. The inclusion of $t_{2g}-t_{2g}$
hoppings will not modify qualitative results, introducing only minor numerical changes. Second, the states with the number of electrons per ion larger than six are unfavorable due to the strong on-site Coulomb repulsion. Third, the transitions of electrons between the
lattice sites corresponding to the changes of spin by more than one half are strongly suppressed since they involve the simultaneous change of a state for two or more electrons.

As a result, in doped cobaltites there remain only two most probable hopping processes: (i) the transitions of electrons between the IS Co$^{3+}$ and LS Co$^{4+}$ and (ii) transitions  between the HS Co$^{3+}$ and IS Co$^{4+}$. The corresponding configurations are illustrated in Table~\ref{Table}. Thus, to facilitate the kinetic energy gain due to the electron transfer, one can create a ground state with intermediate spins. Such a situation can arise if in the ground state for isolated ions, we have either LS Co$^{4+}$ or HS Co$^{3+}$. The former case corresponds to $\Delta > 3J_H$ when some of LS Co$^{3+}$ can be promoted to the IS state. In the latter case corresponding to $\Delta < 2J_H$, some HS Co$^{4+}$ are promoted to the IS state. In the intermediate situation, $2J_H < \Delta < 3J_H$, the electron transfer can occur if we promote both ions, Co$^{3+}$ and Co$^{4+}$, to some excited states. Such double excitations seem to be less probable. Below, we first discuss the most realistic case $\Delta >3J_H$ at different doping levels with a special emphasis on the possibility of phase separation. Then, we perform the similar
study for $\Delta < 2J_H$. After that, we construct the phase
diagram of the system at $\Delta/J_H-$doping plane.

Actually, $\Delta$, or rather $\Delta/t$, regularly depends on the rare earth radius $r_A$ in the series of RCoO$_3$ perovskites and increases with decreasing $r_A$.

\section{Charge transfer effects and spin-state transitions:
The case of LS-LS ground state for isolated ions} \label{LS-LS}

Let us first discuss the situation corresponding to doped perovskite
cobaltites, where Co$^{3+}$ and Co$^{4+}$ ions occupy the sites of a
simple cubic lattice. The relative number of Co$^{4+}$ and Co$^{3+}$
is respectively $x$ and $1-x$. Let us assume that in the absence of
charge transfer the cobalt ions of both types are in the LS state
($\Delta > 3J_H$); this is a typical situation e.g. in the hole-doped La$_{1-x}$Sr$_x$CoO$_3$, and even more so for smaller rare earths R
in doped RCoO$_3$~\cite{Lorenz}. By promoting some Co$^{3+}$ ions to the IS state, we can have a gain in kinetic energy related to the charge transfer from IS Co$^{3+}$ to LS Co$^{4+}$.

To treat this situation in more detail, let us introduce creation
operators $a_{\bf{n}}^{\dag}$ and $c_{\bf{n}}^{\dag}$ for an electron at the $e_g$ level and a hole at the $t_{2g}$ level, respectively, at site $\bf{n}$ according to the following rules (choosing Co$_{LS}^{3+}$ as the vacuum state)
\begin{eqnarray}\label{states}
|0\rangle &=&|\textrm{Co}_{LS}^{3+}\rangle, \quad E^{(vac)} = E_0, \nonumber \\
a_{\bf{n}}^{\dag}|0\rangle &=&|\textrm{Co}^{2+}\rangle, \quad E^{(2+)} = U', \nonumber \\
c_{\bf{n}}^{\dag}|0\rangle &=&|\textrm{Co}_{LS}^{4+}\rangle, \quad E^{(h)} = E_1.
\end{eqnarray}

In terms of these operators, the intermediate-spin state of Co$^{3+}$ ions can be constructed in the following way
\begin{equation}\label{ISstate}
|\textrm{Co}_{IS}^{3+}\rangle = c_{\bf{n}}^{\dag}a_{\bf{n}}^{\dag}|0\rangle,
\quad E_{IS}^{(3+)} = E_0 + \Delta -J_H = E_2.
\end{equation}

Summing up all possible low-energy configurations, we can write the
following single-site Hamiltonian
\begin{eqnarray}\label{H_onsite1}
H_{\bf{n}} &=& E_0 + (E_1-E_0)n_{\bf{n}}^h +(U'-E_0)n_{\bf{n}}^e
+ \nonumber \\
&&\left[(E_2-E_0)-(E_1-E_0)-(U'-E_0)\right]n_{\bf{n}}^hn_{\bf{n}}^e,
\end{eqnarray}
where $n^e_{{\bf n}}= a^{\dag}_{{\bf n}}a_{{\bf n}}$ and $n^h_{{\bf
n}}= c^{\dag}_{{\bf n}}c_{{\bf n}}$ are the operators describing the
numbers of electrons at in $e_g$ levels and holes at $t_{2g}$ levels, respectively. Writing \eqref{H_onsite1} in a more compact form, we have
\begin{eqnarray}\label{H_onsite2}
H_{\bf{n}} &=& [E_0 + (E_1-E_0)(n_{\bf{n}}^h -n_{\bf{n}}^e)]
+ \nonumber \\
&&(\Delta-J_H)n_{\bf{n}}^e +Un_{\bf{n}}^e(1-n_{\bf{n}}^h),
\end{eqnarray}
where $U=U'+E_1-\Delta+J_H$. Taking the sum over all lattice sites
and introducing the intersite hopping terms, we get
\begin{eqnarray}\label{H}
H&=&\sum_{\bf{n}}[E_0 + (E_1-E_0-\mu)(n_{\bf{n}}^h -n_{\bf{n}}^e)]
+ \nonumber\\
&&+\Delta_1\sum_{{\bf n}}n^e_{{\bf n}}+
U\sum_{{\bf n}}n^e_{{\bf n}}(1-n^h_{{\bf n}}) \nonumber\\
&&-t\sum_{\langle{\bf n}{\bf m}\rangle}\left(
a^{\dag}_{{\bf n}}a_{{\bf m}}+h.c.\right)\, ,
\end{eqnarray}
where $\Delta_1 =\Delta - J_H$.

In Hamiltonian~\eqref{H}, we took into account only the most
significant hopping integral $t$ describing the transitions of
electrons from the occupied $e_{g}$ level of IS Co$^{3+}$ to the
empty $e_{g}$ level of LS Co$^{4+}$. Moreover, we assume that an
electron can move only without changing the $z$ projection of its
spin, and, because of  the Hund's rule coupling the spin of an
itinerant ($e_g$) electron and the total spin of core ($t_{2g}$)
electrons are parallel to each other. Hence, we can assume a ferromagnetic ground state and omit a spin index of electron operators. We also neglect the possible complications related to the orbital degeneracy of the $e_{g}$ level occupied by a single electron; these are not crucial for the present problem.

Such simplified Hamiltonian~\eqref{H} is quite similar to that of the Falicov-Kimball model~\cite{fal}. In model~\eqref{H}, we have, in fact, an interplay between the electron localization in the LS state and the itinerancy in the IS state. This kind of interplay was analyzed in detail both analytically~\cite{myPRL} and  numerically~\cite{Ramakr}, and a tendency for a nanoscale phase
separation was demonstrated. The local (atomic scale) charge and spin inhomogeneities related to electronic phase separation were also found recently in exact calculations for small clusters~\cite{Kochar1}. Here, following the technique suggested
in Refs.~\onlinecite{myPRL,myPRB}, we address the specific features
of the systems with the spin-state transitions.

The average numbers of $e_g$ electrons and $t_{2g}$ holes
per site $\langle n^e_{{\bf n}}\rangle = n^e$ and $\langle n^h_{{\bf n}}\rangle = n^h$ obey the evident relationship $n^h - n^e = x$
(by electrons we mean here not the real extra electrons,
which would create the state Co$^{2+}$, but the electrons in the
initially empty $e_g$ levels, promoted there by the LS-IS transition, i.e. we still are dealing with the ``mixture" of Co$^{3+}$ and
Co$^{4+}$ in the hole-doped system).

Then, the energy per site can be written as
\begin{equation}\label{E}
E^{(1)} = E_0(1-x)+ E_1x + \langle H_1 \rangle /N  \, ,
-\end{equation}
where
\begin{eqnarray}\label{H1}
H_1&=& \Delta_1\sum_{{\bf n}}n^e_{{\bf n}}+
U\sum_{{\bf n}}n^e_{{\bf n}}(1-n^h_{{\bf n}}) \nonumber\\
&&-t\sum_{\langle{\bf n}{\bf m}\rangle}\left(
a^{\dag}_{{\bf n}}a_{{\bf m}}+h.c.\right)\, .
\end{eqnarray}

Let us consider a homogeneous state corresponding to a certain
density $n^e$ of electrons promoted to the IS Co$^{3+}$ state. We
calculate the energy spectrum using the Hubbard I decoupling~\cite{HubbardI} in equation of motion for the one-electron Green function $G^e({\bf n,n}_0;\,t-t_0)=-i\langle Ta_{{\bf n}}(t)a^{\dag}_{{\bf n}_0}(t_0)\rangle$ for these promoted $e_g$ electrons (analogous to the band ($b$) electrons in Refs.~\onlinecite{myPRL} and \onlinecite{Ramakr}). In
the frequency-momentum representation
\begin{equation}\label{Gappr} G^e({\bf
k},\omega)=-\frac{\omega+\mu - \Delta_1-Un^h}{\left(\omega+\mu -
\Delta_1-E_1({\bf k})\right)\left(\omega+\mu-\Delta_1-E_2({\bf
k})\right)}\,, \end{equation}
where
\begin{equation}\label{E12}
E_{1,2}({\bf k})=\frac{U+\varepsilon({\bf
k})}{2}\mp\sqrt{\left(\frac{U-\varepsilon({\bf
k})}{2}\right)^2+U\varepsilon({\bf k})(1-n^h)}\,, \end{equation} and
$\varepsilon({\bf k})$ is the energy spectrum at $U=0$. We choose
$\varepsilon({\bf k})$ in the simplest tight-binding form, ignoring
possible orbital effects and not taking into account the specific
features of the hopping integrals of $e_g$ electrons,
[$\varepsilon({\bf k})=-2t(\cos k_x+\cos k_y+\cos k_z)$ for simple
cubic lattice].

Using Eq.~\eqref{Gappr} for the Green function, we calculate the
densities $n^e$ and $n^h$ of $e_g$ electrons and holes. Based on
these results, we can determine the dependence of the total energy
on the doping level $x$. These calculations are similar to those
performed in Ref.~\onlinecite{myPRL}. The $b$ electrons in
Ref.~\onlinecite{myPRL} correspond to our $e_g$ electrons at IS
Co$^{3+}$ ions, whereas the number of localized $l$ electrons in
Ref.~\onlinecite{myPRL}, $n_l$, corresponds to the number of LS
Co$^{3+}$ ions, $1-n^h$. Using this similarity, we could make a
direct mapping between the two systems. However, in the systems
with spin-state transitions there exists another homogeneous state
in addition to that considered in Ref.~\onlinecite{myPRL}. Namely,
it corresponds to all Co$^{3+}$ ions in an intermediate spin state
($n^h=1$, $n^e=1-x$). Formally, in terms of a conduction band and
localized level, this state can be treated as a combination of
empty localized level lying below the Fermi level of the partially
filled conduction band, which is in general not possible. In our
case, the state corresponding to localized level disappears in the
absence of LS Co$^{3+}$ ions.

\begin{figure} \begin{center}
\includegraphics*[width=0.8\columnwidth]{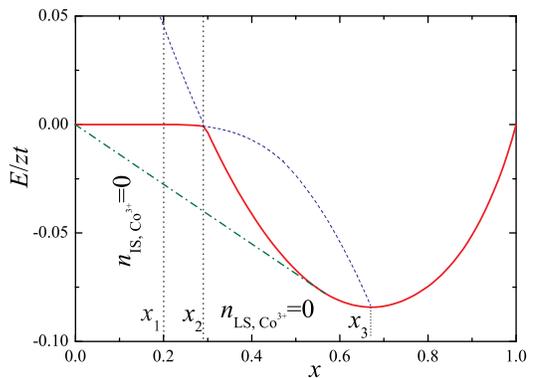}
\end{center} \caption{\label{FigE_LS} (Color online)
The energies of type 1 and 2 states as function of doping $x$.
The red solid curve corresponds to the more favorable in energy
homogeneous state, whereas the states with higher energies are
shown by blue dashed lines (see the text). The green dot-dashed
line corresponds to the energy of an inhomogeneous state,
obtained by Maxwell construction.
$\Delta_1/zt=0.2$.}
\end{figure}

\begin{figure} \begin{center}
\includegraphics*[width=0.8\columnwidth]{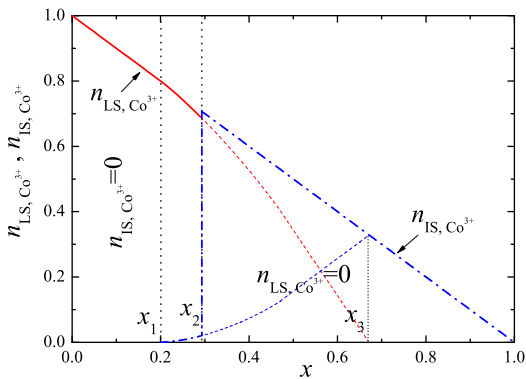}
\end{center} \caption{\label{FigN_LS} (Color online) The densities
of Co$^{3+}$ ions in intermediate- ($n^e=n_{\text{IS, Co}^{3+}}$) and low-spin ($1-n^h=n_{\text{LS, Co}^{3+}}$) homogeneous states as a function of $x$ at $\Delta_1/zt=0.2$. At $x_1 < x < x_2$, both spin states of Co$^{3+}$ coexist, whereas at $x =x_2$, there occurs a jump-like transition to the purely IS state of Co$^{3+}$ ions. Dashed lines illustrate the possible behavior of $n_{\text{IS, Co}^{3+}}$ and $n_{\text{LS, Co}^{3+}}$ for the type I state similar to that
described in Ref.~\onlinecite{myPRL}.} \end{figure}

Let us denote the state similar to that in Ref.~\onlinecite{myPRL}
(that is the state with coexisting LS and IS Co$^{3+}$ ions) as a
type 1 state, and the state without LS Co$^{3+}$ as type 2 state.
The energies of these two states as function of doping $x$ are shown in Fig.~\ref{FigE_LS} at $\Delta_1/zt=0.2$ ($z=6$ is the number of nearest neighbors). The type 2 state becomes favorable at $x>x_2$. Note that at $x>x_3$ both states, 1 and 2, are equivalent. We can see that at $x<x_1$ there are no electrons promoted to the $e_g$ level ($n^e=0$), see Fig.~\ref{FigN_LS}.

At $x>x_1$ the number of $e_g$ electrons gradually grows. In the absence of type 2 state, this growth  would continue up to $x=x_3$ when all Co$^{3+}$ ions would turn to the intermediate spin state. At $x=x_2$, however, the type 2 state becomes favorable in energy, and the jump-like transition to this state occurs. The ground state energy $E$ for the homogeneous system as function of $x$ is shown in Fig.~\ref{FigE_LS} by red solid curve. At the same time, it is
clear from this figure that in the doping range $0<x\lesssim x_3$ the inhomogeneous state, being a mixture of states with $n^h=1$ and $n^e=0$, is more favorable. The energy of this mixed state is shown in Fig.~\ref{FigE_LS} by the green dot-dashed line.

The densities of Co$^{3+}$ ions in intermediate-spin ($n^e$) and
low-spin ($1-n^h$) states as a function of $x$ are shown in Fig.~\ref{FigN_LS} by blue dot-dashed and red solid curves, respectively. We see a jump-like increase of $n^e$ at $x_2$,
when the homogeneous type 2 state would become favorable. The behavior of $n^e$ and $1-n^h$ at $x>x_2$ in the absence of type 2 state are shown by thin blue and red dashed lines, respectively.

At a small band filling, $n^e\ll 1$, we can write an approximate
explicit expression for the total energy $E$ assuming that the
Fermi surface is spherical
\begin{equation}\label{Etot0} E\simeq
\Delta_1n^e-tzn^en^h+
\frac{3t}{5}\left(36\pi^4n^h\right)^{1/3}(n^e)^{5/3}\,.
\end{equation}
The density $n^e$ of itinerant electrons is determined by
minimization of  Eq.~\eqref{Etot0} with respect to $n^e$ taking into
account that $n^h = x + n^e$. It can be easily shown that the
solution for the energy minimum corresponding to $n^e\neq 0$ can
exist only if $\Delta_1 < tzx$. This means that at $\Delta_1/tz > 1$
the LS Co$^{3+}$ ions can not be promoted to the IS state at any
doping $x$.

\begin{figure}[!hbt]\centering
   \subfigure[]{
      \includegraphics[width=0.8\columnwidth]{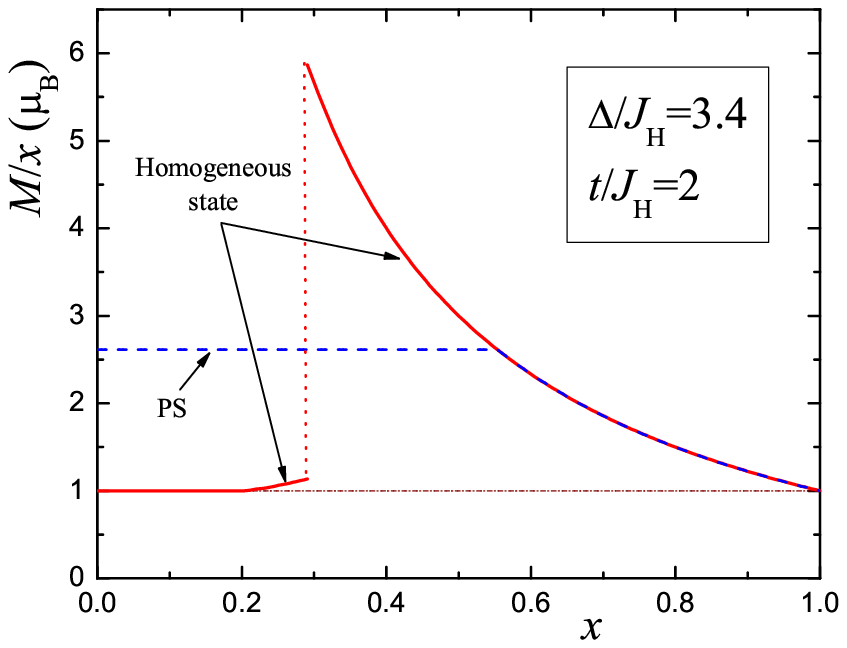}}
   \subfigure[]{
      \includegraphics[width=0.8\columnwidth]{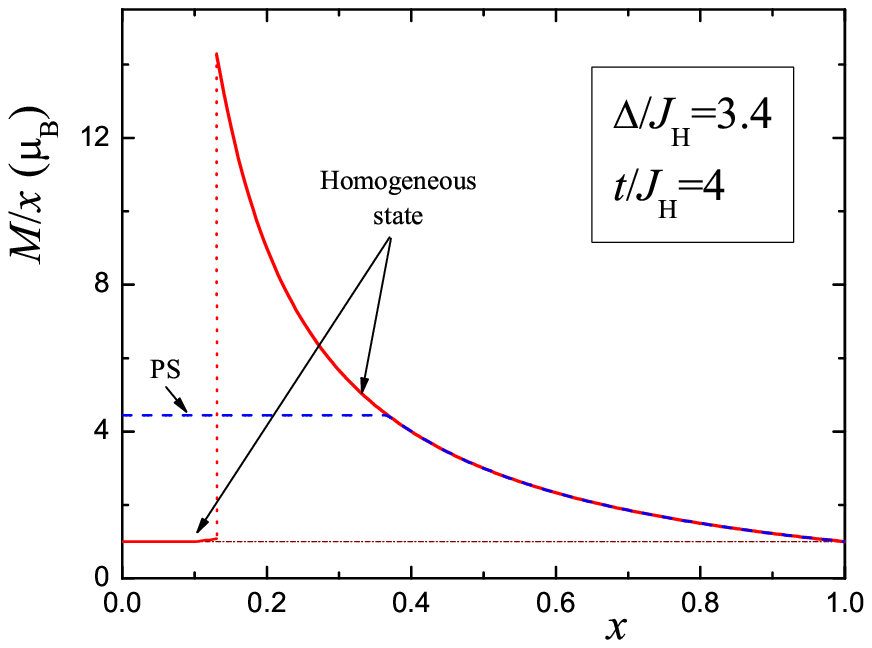}}
   \caption{(Color online) Magnetic moment per one Co$^{4+}$
ion versus doping $x$ at different values of the hopping
integral $t$. The red curves correspond to the homogeneous
states. The behavior $M$ in the phase-separated state
is shown by the blue dashed line.}
\label{FigM_xLS} \end{figure}

The dependence of $n^e$ on doping $x$ determines the behavior of
magnetic moment of Co ions. Indeed, the LS Co$^{3+}$ ions correspond
to zero magnetic moment, $S=0$, while the doping leads to creation
of LS Co$^{4+}$ ions ($S=1/2$) and also provides the promotion of
some Co$^{3+}$ ions to the IS state ($S=1$). So, the data presented
in Fig.~\ref{FigN_LS} could be redrawn in terms of magnetic moment per dopant (or, in other words, per Co$^{4+}$), see Fig.~\ref{FigM_xLS}. For the homogeneous state, we see in Fig.~\ref{FigM_xLS} that the jump-like transition in the density $n^e$ of itinerant electrons manifests itself in a jump of magnetic moment. At the same time, in the phase-separated state, the magnetic moment per Co$^{4+}$ ion remains constant since  both the content of the phase with IS Co$^{3+}$ and the number of Co$^{4+}$ ions are proportional to $x$. The value of magnetic moment per Co$^{4+}$ is determined by the value of $x$, where the green dot-dashed line in Fig.~\ref{FigE_LS} touches the curve corresponding to the energy of the homogeneous state. Both the height of the jump for the magnetic moment in the homogeneous state and the value of magnetic moment per Co$^{4+}$ in the phase-separated state depend drastically in the parameters of the model, especially on the hopping integral $t$. In Fig.~\ref{FigM_xLS}, we see that the increase in $t$ by a factor of two leads to a pronounced growth of both mentioned values. Note here that the values of magnetic moment under discussion correspond to macroscopic phase separation, that is the characteristic sizes of
inhomogeneities are much larger than the lattice constant. It is
indeed so at relatively large $x$ (exceeding the percolation
threshold for the phase with itinerant charge carriers). At small
$x$, it is naturally to expect that the phase-separated system
will consist of small droplets (spin-state polarons) containing
only one Co$^{4+}$ ion surrounded by IS Co$^{3+}$. In the latter
case, the magnetic moment per Co$^{4+}$ should be larger than
that corresponding to the macroscopic phase separation. This
could be the case for spin polarons in low-doped
La$_{1-x}$Sr$_x$CoO$_3$ observed in Ref.~\onlinecite{Podles08},
where the polarons with the magnetic moment equal to $13\mu_B$
seem to be the most probable. From our considerations, one should expect that the value of the moment per of Co$^{4+}$ should become smaller with the increase of doping $x$.
The exact calculations for small clusters also demonstrate that in a suitable range of parameters the saturated magnetic moment can exist at relatively low temperatures also in atomic-size doped clusters of various geometries~\cite{Kochar2}.

Thus, we demonstrated that the spin-state transitions in hole-doped
cobaltites can be described based on the model involving the
coexistence and competition of localized and itinerant electron
states. In contrast to the similar model for
manganites~\cite{myPRL,myPRB}, this model
allows the possibility of a jump-like transition to the purely
itinerant state corresponding in the case of cobaltites to the LS $\rightarrow$IS transition for all Co$^{3+}$ ions. However, at lower doping, before reaching this homogeneous metallic state with all
Co ions magnetic, the phase-separated state comes into play, in which only a part of Co$^{3+}$ ions is promoted to the IS state, doped holes being located in these regions. Experimental data on La$_{1-x}$Sr$_x$CoO$_3$~\cite{Rivas,Loshk,Louca,Leighton06,
Leighton07,Podles08} seem to be in agreement with this picture.

\section{Charge transfer effects and spin-state transitions:
The case of HS-HS ground state for isolated ions} \label{HS-HS}

Let us now discuss the situation at $\Delta < 2J_H$, when in the
absence of electron hopping it is favorable for both Co$^{3+}$ and
Co$^{4+}$ to be in the HS state. The charge transfer becomes
possible only if we promote a hole to the $e_g$ level of Co$^{4+}$ and transform such an ion from HS to IS state.

So, in this case, instead of electron hopping from IS Co$^{3+}$ to LS Co$^{4+}$, we have the electron hopping from the HS Co$^{3+}$ to IS
Co$^{4+}$, or the hole hopping from IS Co$^{4+}$ to HS Co$^{3+}$ (this representation is more convenient here). Using this analogy, we can choose the HS state of Co$^{4+}$ as a new vacuum state and write relationships similar to \eqref{states} and \eqref{ISstate} as
\begin{eqnarray}\label{statesHS}
|0\rangle &=&|\textrm{Co}_{HS}^{4+}\rangle, \quad \tilde{E}^{(vac)} = E_1 +2\Delta -6J_H =\tilde{E}_0,
\nonumber \\
\tilde{c}_{\bf{n}}^{\dag}|0\rangle &=&|\textrm{Co}^{5+}\rangle, \quad E^{(5+)} = \tilde{U}',
\nonumber \\
\tilde{a}_{\bf{n}}^{\dag}|0\rangle &=&|\textrm{Co}_{HS}^{3+}\rangle,
\quad E^{(e)} = E_0+2\Delta-4J_H = \tilde{E}_1, \nonumber \\
\tilde{a}_{\bf{n}}^{\dag}\tilde{c}_{\bf{n}}^{\dag}|0\rangle &=&|\textrm{Co}_{IS}^{4+}\rangle,
\, E_{IS}^{(4+)} = E_0 + \Delta -2J_H = \tilde{E}_2.
\end{eqnarray}

The corresponding single-site Hamiltonian can be found by the
following substitution in \eqref{H_onsite1} and \eqref{H_onsite2}:
$E_0, E_1, E_2, U \rightarrow \tilde{E}_0, \tilde{E}_1, \tilde{E}_2,
\tilde{U}$ and also $n^e_{{\bf n}} \rightarrow \tilde{n}^h_{{\bf n}},
n^h_{{\bf n}} \rightarrow \tilde{n}^e_{{\bf n}}$.

As a result, we can rewrite the Hamiltonian~\eqref{H} in the
following form
 \begin{eqnarray}\label{H_HS}
H&=&\sum_{\bf{n}}[\tilde{E}_0 + (\tilde{E}_0-\tilde{E}_1-\mu)
(\tilde{n}_{\bf{n}}^e -\tilde{n}_{\bf{n}}^h)]
+ \nonumber\\
&&+\Delta_2\sum_{{\bf n}}\tilde{n}^h_{{\bf n}}+
\tilde{U}\sum_{{\bf n}}\tilde{n}^h_{{\bf n}}(1-\tilde{n}^n_{{\bf n}}) \nonumber\\
&&-t\sum_{\langle{\bf n}{\bf m}\rangle}\left(
\tilde{c}^{\dag}_{{\bf n}}\tilde{c}_{{\bf m}}+h.c.\right)\, .
\end{eqnarray}

Here, $\Delta_2 = 4J_H -\Delta$ is the energy
difference between the IS and HS Co$^{4+}$ ions,
$\tilde{c}^{\dag}_{{\bf n}}$, $\tilde{c}_{{\bf n}}$ are creation
and annihilation operators for a hole promoted to the $e_{g}$
level of IS Co$^{4+}$ at site $\bf n$, $\tilde{n}^h_{{\bf n}}=
\tilde{c}^{\dag}_{{\bf n}}\tilde{c}_{{\bf n}}$, and
$\tilde{n}^e_{{\bf n}} = \tilde{a}^{\dag}_{{\bf n}}\tilde{a}_{{\bf
n}}$ is the operator describing the number ($0$ or $1$) of
additional localized $t_{2g}$ electrons at site ${\bf n}$
($\tilde{a}^{\dag}_{{\bf n}}$, $\tilde{a}_{{\bf n}}$ are creation
and annihilation operators for such electrons). The average
numbers of electrons and holes per site obey now the relationship
$\tilde{n}^e - \tilde{n}^h = 1-x$.

In this case, the energy per site \eqref{E} can be rewritten as
\begin{equation}\label{E2}
E^{(2)} = E_0(1-x)+ E_1x + \langle H_2 \rangle /N  \, ,
\end{equation}
where
\begin{eqnarray}\label{H1a}
H_2&=& \sum_{{\bf n}}\left(2\Delta - 6J_H +
2J_H(\tilde{n}_{\bf{n}}^e -
\tilde{n}_{\bf{n}}^h)\right)+ \nonumber\\
&&\Delta_2\sum_{{\bf n}}\tilde{n}_{\bf{n}}^h  +
\tilde{U}\sum_{{\bf n}}\tilde{n}^h_{{\bf n}}
(1-\tilde{n}^n_{{\bf n}})\nonumber\\
&&-t\sum_{\langle{\bf n}{\bf m}\rangle}\left(
\tilde{a}^{\dag}_{{\bf n}}\tilde{a}_{{\bf m}}+h.c.\right)\, .
\end{eqnarray}
Note that the difference $E^{(2)} - E^{(1)}$, does not depend on
the choice of $E_0$ and $E_1$, this fact will be helpful in
constructing the phase diagrams in the next section.

Thus, the behavior of the system energy and charge carrier
densities, $\tilde{n}^e$ and $\tilde{n}^h$ are similar to those
shown in Figs.~\ref{FigE_LS} and~\ref{FigN_LS}. In  these figures
we should replace $n^h\to\tilde{n}^e$, $n^e\to\tilde{n}^h$, and
$x \to 1-x$, that is, the densities of Co$^{3+}$ ions in IS
($n^e=n_{\text{IS,Co}^{3+}}$) and LS ($1-n^h=n_{\text{LS, Co}^{3+}}$) states become  here the densities of Co$^{4+}$ ions in IS
($\tilde{n}^h=n_{\text{IS, Co}^{4+}}$) and HS
($1-\tilde{n}^e=n_{\text{HS, Co}^{4+}}$) states.
Note also that such an exact similarity between the LS-LS and HS-HS cases appears since we, in fact, deal with the spinless fermions (the spins of charge carriers are parallel). So, we have an electron-hole symmetry between an empty $e_g$ level at LS Co$^{3+}$ and a completely occupied such level at HS Co$^{3+}$.

\section{Phase diagrams}

\begin{figure}[!hbt]\centering
   \subfigure[$\quad t/J_H = 1$]{
      \includegraphics[width=0.9\columnwidth]{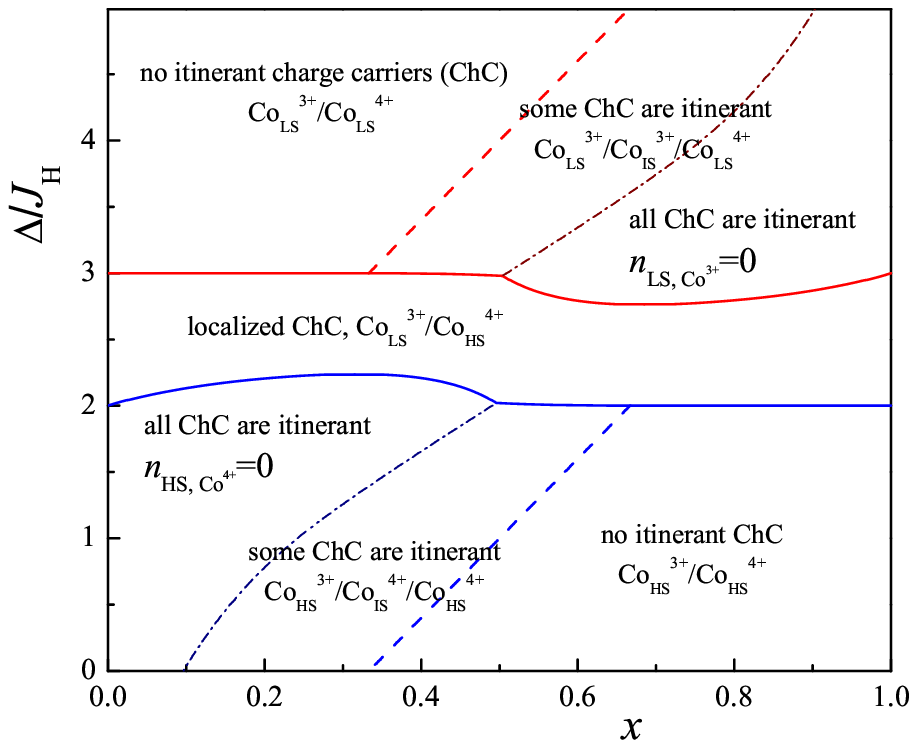}}
   \subfigure[$\quad t/J_H = 1.5$]{
      \includegraphics[width=0.9\columnwidth]{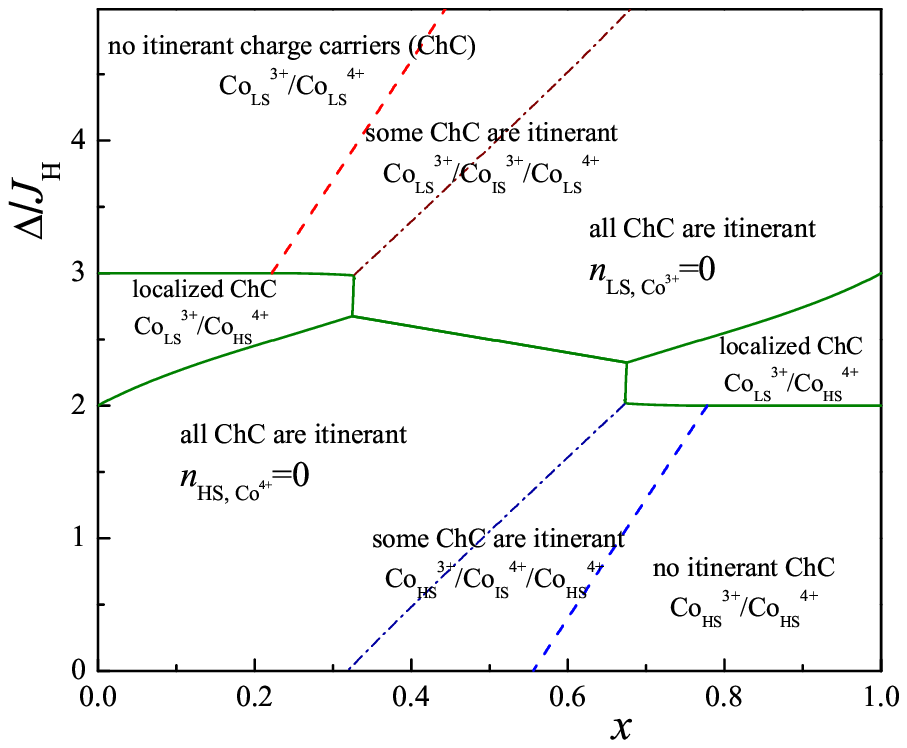}}
   \caption{(Color online) Possible homogeneous states of the system
under study at different values of the hopping integral $t$. The boundaries between analogous phases in the upper and lower parts of the phase diagram are shown by the same lines (solid, dashed, or dot-and-dash).}
\label{PhDiaHom} \end{figure}

Based on the results of the previous sections, we can summarize the
behavior of the system as function of doping at different values of
the $\Delta/J_H$ ratio and to draw the corresponding phase diagram.
The form of this phase diagram depends drastically on the
characteristic values of the hopping integral $t$. The general
features of the evolution of the system with doping from one
homogeneous state to another are illustrated in Fig.~\ref{PhDiaHom}.
At rather small $t$ ($t/J_H \lesssim 1$), see Fig.~\ref{PhDiaHom}a,
we have clearly defined regions of the phase diagram corresponding
to $\Delta > 3J_H$ and $\Delta < 2J_H$ (corresponding to the
situations discussed in sections \ref{LS-LS} and \ref{HS-HS},
respectively). In each of these regions, the variation of doping
leads to the transitions between the phase with only localized
carriers to the phase when some charge carriers are delocalized
and, eventually, to the phase when all charge carriers are itinerant.
These two regions, with $\Delta > 3J_H$ and $\Delta < 2J_H$, are
separated by the phase with Co$^{3+}$ in LS ($S=0$) and Co$^{4+}$ in
HS ($S=5/2$) states,with the charge carriers localized because of
the spin blockade~\cite{SpinBlock}.
At larger  $t$ ($t/J_H \gtrsim 1$), see Fig.~\ref{PhDiaHom}b, the
latter intermediate region collapses at a certain doping range, and
a direct spin-state transition between the phases with fully
delocalized charge carriers becomes possible.

\begin{figure}\centering
   \subfigure[$\quad t/J_H = 1$]{
      \includegraphics[width=0.9\columnwidth]{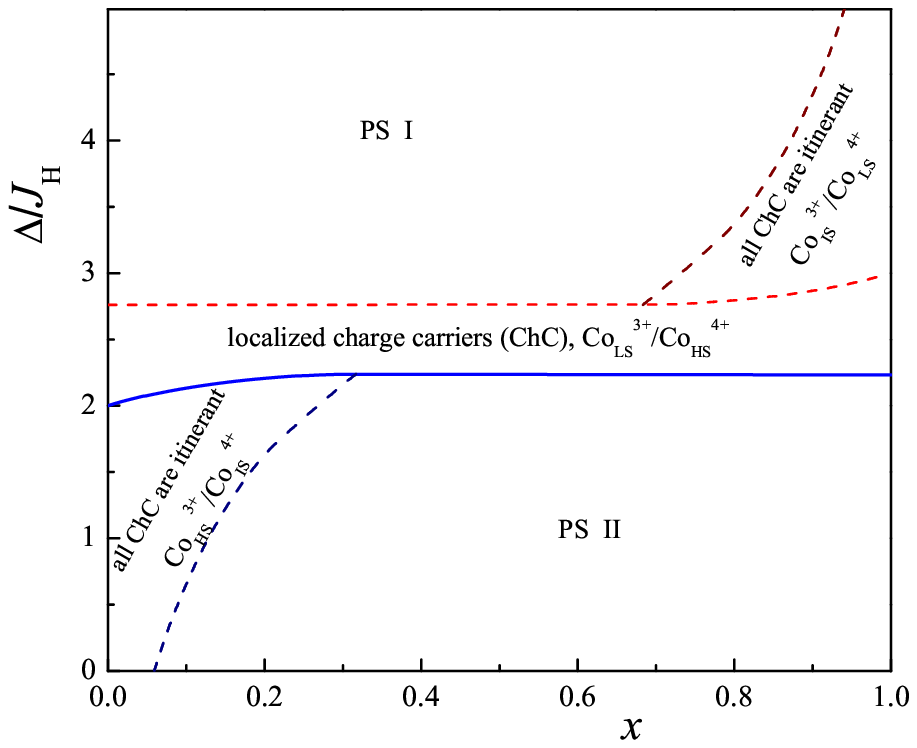}}
    \subfigure[$\quad t/J_H = 1.3$]{
      \includegraphics[width=0.9\columnwidth]{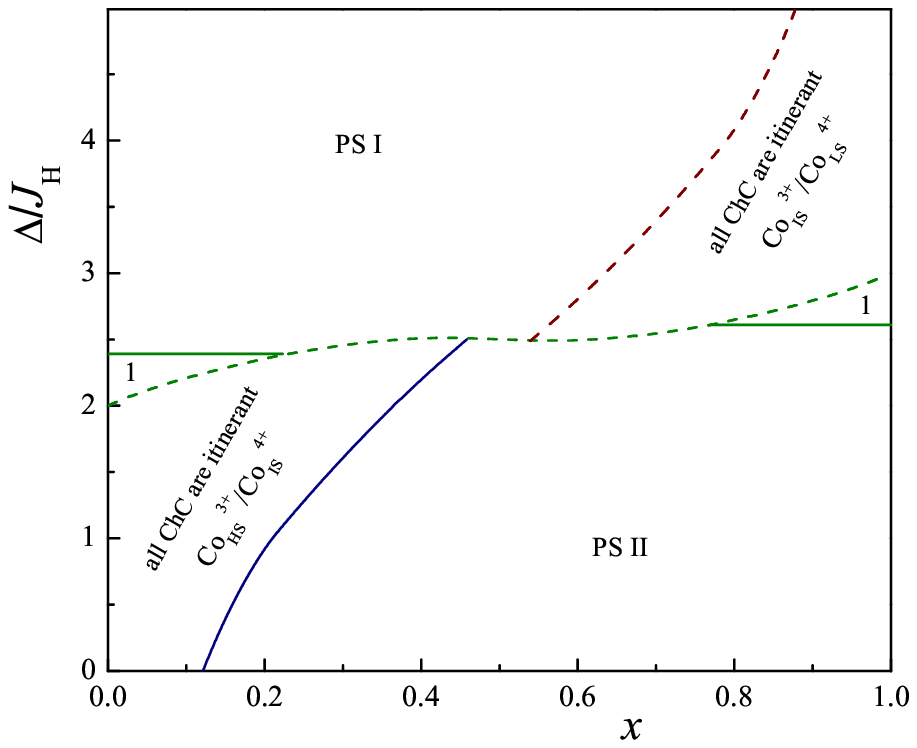}}
   \subfigure[$\quad t/J_H = 1.5$]{
      \includegraphics[width=0.9\columnwidth]{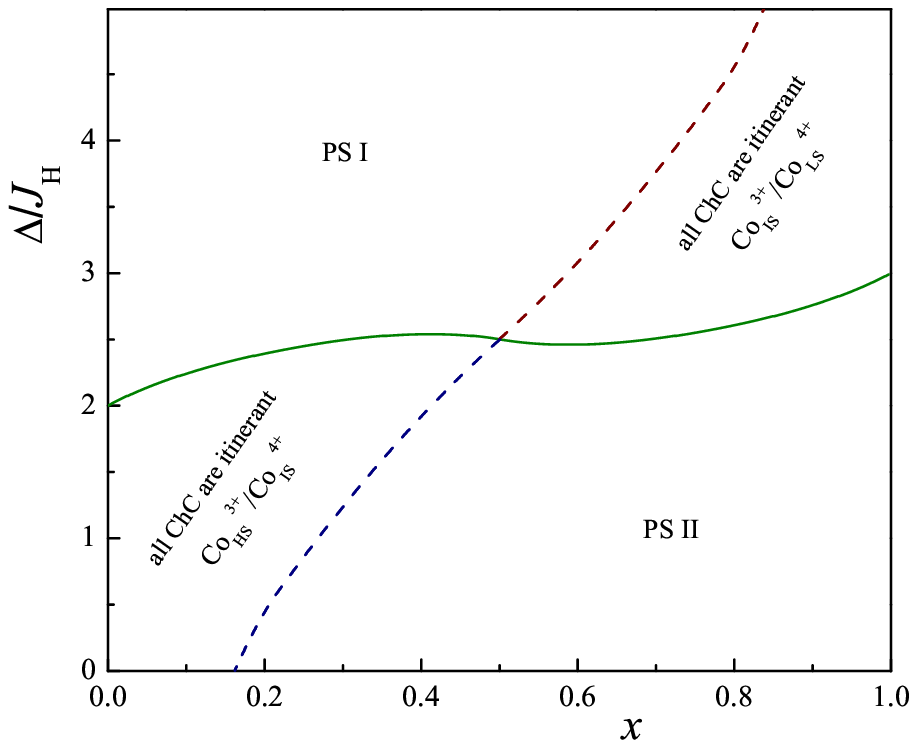}}
   \caption{(Color online) Phase diagrams including the phase-separated states of the system under study at different values of hopping integral $t$. PS I is the phase-separated state including the regions without itinerant charge carriers, corresponding to LS Co$^{3+}$, and those with completely delocalized charge carries promoted to IS Co$^{3+}$. PS II is the similar phase-separated state where the regions with  and without itinerant charge carriers correspond to IS Co$^{4+}$ and HS Co$^{4+}$, respectively. Regions 1 at panel (b) correspond to the charge carriers located at HS Co$^{4+}$ and LS Co$^{3+}$ where the charge transfer between Co sites is suppressed due to the spin blockade.}
\label{PhDiaInhom} \end{figure}

The form of the phase diagram changes if we take into account the
possibility of phase separation. The corresponding phase diagrams
drawn at different values of $t/J_H$ are shown in Fig.~\ref{PhDiaInhom}. We see that instead of phases with partially
and fully delocalized charge carriers, there appears a broad regions
of phase separation where the domains of fully localized and fully
delocalized charge carriers are intermixed. Again, at rather small
$t$ ($t/J_H \lesssim1$), we have and intermediate region where  the
charge carries are localized at any doping level (with Co$^{3+}$ and
Co$^{4+}$ in LS and HS states, respectively). This intermediate
region gradually disappears with the growth of the hopping integral
$t$.

Let us note here that the phase diagram along the
$\Delta$ axis could be reproduced varying the average ionic
radius of the rare-earth ions in cobaltites (see, e.g.
Refs.~\onlinecite{Fujita}, \onlinecite{Wang}).

Note also that the long-range Coulomb interaction related to the charge disproportionalization in the phase-separated state can reduce the doping range of the phase separation and modify the form of the phase diagram shown in Fig.~\ref{PhDiaInhom}.

\section{Conclusions}

Based on a simplified model of a strongly correlated
electron system with spin-state transitions, we demonstrated a
tendency to the phase separation for doped perovskite cobaltites in
a wide range of doping levels. The phase diagram including large regions of inhomogeneous phase-separated states was constructed in the plane of parameters doping $x$ versus $e_g-t_{2g}$ energy splitting $\Delta$. The form of the phase diagram turns out to be strongly dependent on the ratio of of the electron hopping integral $t$ and and the Hund's rule coupling constant $J_H$.

Here, we did not analyzed in detail
the possible structure of the phase-separated state. However, for the corresponding model describing doped manganites, the calculations~\cite{myPRL} and numerical simulations~\cite{Ramakr} taking into account the surface and long-range Coulomb contributions to the total energy lead to the characteristic size of nanoscale inhomogeneities of the order of several lattice constants.

\section*{Acknowledgments}

The work was supported by the European project CoMePhS, International Science and Technology Center (grant G1335), Russian Foundation for Basic Research (projects 07-02-91567 and 08-02-00212), and by the Deutsche Forschungsgemeinshaft via SFB 608 and the German-Russian project 436 RUS 113/942/0. A.O.S. also acknowledges support from the
Russian Science Support Foundation.


\begin{thebibliography} {99}

\bibitem[\S]{affUK}
Also at the Department of Physics, Loughborough University,
Leicestershire, LE11 3TU, UK.

\bibitem{myPRL}
K.I. Kugel, A.L. Rakhmanov, and A.O. Sboychakov, \prl {\bf 95},
267210 (2005).

\bibitem{dagbook}
E. Dagotto, \textit{Nanoscale Phase Separation and Colossal
Magnetoresistance: The Physics of Manganites and Related Compounds}
(Springer-Verlag, Berlin, 2003).

\bibitem{Nag} E. Nagaev, \textit{Colossal Magnetoresistance
and Phase Separation in Magnetic Semiconductors} (Imperial
College Press, London, 2002).

\bibitem{Kak}
M.Yu. Kagan and K.I. Kugel, Usp. Fiz. Nauk. {\bf 171}, 577 (2001)
[Physics - Uspekhi {\bf 44}, 553 (2001)].

\bibitem{orbPS1}
K.I. Kugel, A.L. Rakhmanov, A.O. Sboychakov,
and D.I. Khomskii, \prb {\bf 78}, 155113 (2008).

\bibitem{orbPS2}
K.I. Kugel, A.O. Sboychakov, and D.I. Khomskii,
J. Supercond. Nov. Magn. {\bf 22}, 147 (2009).

\bibitem{Gooden67}
P.M. Raccah and G.B. Goodenough, Phys. Rev. {\bf 155}, 932 (1967).

\bibitem{Tokura1}
S. Yamaguchi, Y. Okimoto, H. Taniguchi, and Y. Tokura, \prb {\bf 53},
R2926 (1996).

\bibitem{Tokura2}
S. Yamaguchi, Y. Okimoto, and Y. Tokura, \prb {\bf 55}, R8666 (1997).

\bibitem{Korotin}
M.A. Korotin,  S.Yu. Ezhov, I.V. Solovyev, V.I. Anisimov, D.I. Khomskii, and G.A. Sawatzky, \prb {\bf 54}, 5309 (1996).

\bibitem{Doumerc}
M. Pouchard, A. Villesuzanne, and J.P. Doumerc, J. Solid State Chem.
{\bf 162}, 282 (2001).

\bibitem{KhoLow}
D.I. Khomskii and U. L\"{o}w, \prb {\bf 69}, 184401 (2004).

\bibitem{AndHas}
P.W. Anderson and H. Hasegawa, Phys. Rev. {\bf 100} 675 (1955).

\bibitem{deGennes}
P.G. de Gennes,  Phys. Rev. {\bf 118} 141 (1960).

\bibitem{BulKhom67}
L.N. Bulaevskii and D.I. Khomskii, Zh. Eksp. Teor. Fiz. {\bf 52},
1603 (1967) [Sov. Phys. JETP {\bf 25}, 1067 (1967)].

\bibitem{BulNagKhom}
L.N. Bulaevskii, E.L. Nagaev, and  D.I. Khomskii, Zh. Eksp. Teor.
Fiz. {\bf 54}, 1562 (1968) [Sov. Phys. JETP {\bf 27}, 836 (1968)].

\bibitem{Nag67}
E.L. Nagaev, Pis'ma Zh. Eksp. Teor. Fiz. {\bf 6}, 484 (1967) [JETP
Lett. {\bf 6}, 18 (1967)].

\bibitem{Kasuya}
T. Kasuya, A. Yanase, and T. Takeda, Solid State Commun. {\bf 8},
1543 (1970).

\bibitem{KilKha}
R. Kilian and G. Khaliullin, \prb {\bf 60}, 13458 (1999).

\bibitem{MiKhoSaw}
T. Mizokawa, D.I. Khomskii, and G.A. Sawatzky, \prb {\bf 61}, R3776 (2000);
\prb {\bf 63}, 024403 (2001).

\bibitem{SpinBlock}
A. Maignan, V. Caignaert, B. Raveau, D. Khomskii, and G. Sawatzky,
\prl {\bf 93}, 026401 (2004).

\bibitem{Rivas}
R. Caciuffo, D. Rinaldi, G. Barucca, J. Mira, J. Rivas, M.A.
Se\~nar\'is-Rodr\'iguez, P.G. Radaelli, D. Fiorani, and J.B. Goodenough, \prb {\bf 59}, 1068 (1999).

\bibitem{Loshk}
N.N. Loshkareva, E.A. Gan'shina, B.I. Belevtsev, Yu.P. Sukhorukov,
E.V. Mostovshchikova, A.N. Vinogradov, V.B. Krasovitsky, and I.N.
Chukanova, \prb {\bf 68}, 024413 (2003).

\bibitem{Louca}
D. Phelan, Despina Louca, K. Kamazawa,  S.-H. Lee, S. Rosenkranz, Y. Motome, M.F. Hundley, J.F.  Mitchell, S.N. Ancona, and Y. Moritomo,
\prl {\bf 97}, 235501 (2006).

\bibitem{Leighton06}
S.R. Giblin, I. Terry, D. Prabhakaran, A.T. Boothroyd, J. Wu, and C.
Leighton, \prb {\bf 74}, 104411 (2006).

\bibitem{Leighton07}
C. He, M.A. Torija, J. Wu, J.W. Lynn, H. Zheng, J.F. Mitchell,
and C. Leighton, \prb {\bf 76}, 014401 (2007).

\bibitem{Podles08}
A. Podlesnyak, M. Russina, A. Furrer, A. Alfonsov, E. Vavilova,
V. Kataev, B. B\"{u}chner, Th. Str\"{a}ssle, E. Pomjakushina,
K. Conder, and D.I. Khomskii, \prl {\bf 101}, 247603 (2008).

\bibitem{KaKhoMo}
M.Yu. Kagan, D.I. Khomskii, and M.V. Mostovoy, Eur. Phys. J. B {\bf
12}, 217 (1999).

\bibitem{KaKuKh01}
M.Yu. Kagan, K.I. Kugel, and D.I. Khomskii, Zh. Eksp. Teor. Fiz.
{\bf 120}, 470 (2001) [JETP {\bf 93}, 415 (2001)].

\bibitem{myPRB}
A.O. Sboychakov, K.I. Kugel,and A.L. Rakhmanov, \prb {\bf 74},
014401 (2006).

\bibitem{Lorenz}
J. Baier, S. Jodlauk, M. Kriener, A. Reichl, C. Zobel,
H. Kierspel, A. Freimuth, and T. Lorenz, \prb {\bf 71},
014443 (2005).

\bibitem{fal}
L.M.~Falicov and J.C.~Kimball, \prl {\bf 22}, 997 (1969).

\bibitem{Ramakr}
V.B. Shenoy, T. Gupta, H.R. Krishnamurthy, and T.V. Ramakrishnan,
\prl {\bf 98}, 097201 (2007).

\bibitem{Kochar1}
A.N. Kocharian, G.W. Fernando, K. Palandage, and J.W. Davenport, Phys. Lett. A {\bf 373}, 1074 (2009).

\bibitem{HubbardI}
J. Hubbard, Proc. Roy. Soc. (London) {\bf A276}, 238 (1963).

\bibitem{Kochar2}
A.N. Kocharian, G.W. Fernando, K. Palandage, and J.W. Davenport, \prb {\bf 78}, 075431 (2008).

\bibitem{Fujita}
T. Fujita, S. Kawabata, M. Sato, N. Kurita, M. Hedo, and Y. Uwatoko,
J. Phys. Soc. Japan {\bf 74}, 2294 (2005).

\bibitem{Wang}
G.Y. Wang, X.H. Chen, T. Wu, G. Wu, X.G. Luo, and C.H. Wang, \prb
{\bf 74}, 165113 (2006).

\end{thebibliography}
\end{document}